\documentclass[aps,pra,amssymb,twocolumn,superscriptaddress,showpacs]{revtex4}

\usepackage{graphicx}
\usepackage{dcolumn}
\usepackage{bm}
\usepackage{color}

\begin{document}
\voffset1cm

\newcommand{\beq}{\begin{equation}}
\newcommand{\eeq}{\end{equation}}
\newcommand{\barr}{\begin{eqnarray}}
\newcommand{\earr}{\end{eqnarray}}

\newcommand{\REV}[1]{\textbf{\color{red}#1}}
\newcommand{\BLUE}[1]{\textbf{\color{blue}#1}}
\newcommand{\GREEN}[1]{\textbf{\color{green}#1}}

\newcommand{\andy}[1]{ }

\newcommand{\bmsub}[1]{\mbox{\boldmath\scriptsize $#1$}}

\def\R{\mathbb{R}}

\def\bra#1{\langle #1 |}
\def\ket#1{| #1 \rangle}
\def\sinc{\mathop{\text{sinc}}\nolimits}
\def\cV{\mathcal{V}}
\def\cH{\mathcal{H}}
\def\cT{\mathcal{T}}
\renewcommand{\Re}{\mathop{\text{Re}}\nolimits}
\newcommand{\tr}{\mathop{\text{Tr}}\nolimits}

\title{
Radon transform on the cylinder and tomography of a particle on the
circle}

\author{M. Asorey}
\affiliation{Departamento de F\'\i sica Te\'orica, Facultad de
Ciencias, Universidad de Zaragoza, 50009 Zaragoza, Spain}
\author{P. Facchi}
\affiliation{Dipartimento di Matematica, Universit\`a di Bari,
        I-70125  Bari, Italy}
\affiliation{INFN, Sezione di Bari, I-70126 Bari, Italy}
\author{V.I. Man'ko}
\affiliation{P.N. Lebedev Physical Institute, Leninskii Prospect
53, Moscow 119991, Russia}
\author{G. Marmo} \affiliation{Dipartimento di Scienze Fisiche,
Universit\`a di Napoli ``Federico II", I-80126  Napoli, Italy}
\affiliation{INFN, Sezione di Napoli, I-80126  Napoli, Italy}
\author{S. Pascazio} \affiliation{Dipartimento di Fisica,
Universit\`a di Bari,
        I-70126  Bari, Italy}
\affiliation{INFN, Sezione di Bari, I-70126 Bari, Italy}
\author{E.G.C. Sudarshan} \affiliation{Department of Physics,
University of Texas, Austin, Texas 78712, USA}

\date{\today}

\begin{abstract}
The tomographic probability distribution on the phase space
(cylinder) related to a circle or an interval is introduced. The
explicit relations of the tomographic probability densities and the
probability densities on the phase space for the particle motion on
a torus are obtained and the relation of the suggested map to the
Radon transform on the plane is elucidated. The generalization to
the case of a multidimensional torus is elaborated and the
geometrical meaning of the tomographic probability densities as
marginal distributions on the helix discussed.
\end{abstract}


\maketitle

\section{Introduction}
\label{sec-introd}

The Radon transform \cite{Rad1917} is the key mathematical tool to
reconstruct the tomographic map of both the Wigner quasidistribution
\cite{Wig32,Moyal,Hillary84} of a quantum state
\cite{Ber-Ber,Vog-Ris,Mancini95} and the probability distribution
on the phase space of a classical particle
\cite{Olga97,ManMenPhysD}.
In the quantum case, this subject not only motivated refined
theoretical approaches based on the maximum likelihood estimation,
in order to extract the maximum reliable information
\cite{theory}, but also interesting experiments with photonic states
\cite{SBRF93}, photon number distributions \cite{torino} and
(helium) atoms \cite{konst}, focusing in particular on the
reconstruction of the transversal motional states. A scheme has been
also proposed in order to obtain the tomographic map associated with
the longitudinal motion of a neutron wave packet
\cite{reconstruct06}. Recent progress on the quantum aspects has
been driven by modern experimental techniques and good reviews on
these topics can be found in \cite{Jardabook}.

The tomographic map provides the symplectic tomography
\cite{dariano96} of quantum states connected with the symplectic
transform on the phase space (the plane $\mathbb{R}^2$ for one
degree of freedom) and this map can be considered as a specific
tomographic version of the star-product quantization
\cite{MarmoJPA,MarmoPhysScr}. Notice that this interpretation of the
Radon transform differs from the original motivation for the Radon
transform in a essential way. The genuine Radon transform  was
introduced as an integral transform defined over submanifolds of the
configuration space, more specifically geodesics (i.e., straight
lines in $\mathbb{R}^2$), whereas in symplectic tomography it is
rather associated to Lagrangian submanifolds of phase space.
Therefore, although  we consider motion, this is instrumental for
the identification of the relevant phase space, but the actual
motions (the solutions of the associated Hamilton equations) do not
appear in the definition of the Radon transform.

If we consider the classical motion of a particle on a circle and
its trajectory in  phase space (a cylinder of radius $R$), the
motion is described by the time dependence of the coordinate
$q(t)=R\phi (t)$, where $\phi (t)$ is the angle defining the point
on the circle. The angular momentum $J$ is the longitudinal
coordinate of this motion in the phase space. In the presence of
fluctuations, the particle state is not determined by the two
coordinates $q$ and $p$ (or $\phi$ and $J$), but rather by their
probability distribution function $f(q,p)$ (or $f(\phi,J)$) on the
phase space. The invertible tomographic map of this distribution
onto the tomographic probability distribution enables one to
determine the state of the classical particle by means of the
probability density $\omega_f(X,\mu,\nu)$, that depends on a random
variable $X$ and two parameters $\mu$ and $\nu$. The parameters
$\mu$ and $\nu$ label the reference frame in the phase space, when
the random position $X$ of the particle is measured. The reference
frame is obtained from the initial one by first squeezing the axis
$q\rightarrow q'=sq$, $p\rightarrow p'=s^{-1}p$, and then performing
the rotation $q'\rightarrow q''=q'\cos\theta+p'\sin\theta$,
$p'\rightarrow p''=-q'\sin\theta+p'\cos\theta$ (see formulae below).
Thus the real parameters $\mu$ and $\nu$ are expressed in terms of
the squeezing $s$ and rotation  $\theta$  as $\mu=s\cos\theta$ and
$\nu=s^{-1}\sin\theta$. The tomographic Radon transform maps the
probability density, that depends on two random variables---position
and momentum---onto the tomographic probability distribution of only
one random variable.

The case of the motion on the circle can be viewed in the limiting
case $R \to \infty$ as the motion on the line. Since the tomographic
map for the classical motion on the line is known (and it is very
similar to the standard Radon transform), it is interesting to
address the question of whether it is possible to describe the
classical motion on the circle by an analogous probability density
distribution depending on one random variable and some extra
parameters. The motion that we consider is purely instrumental in
order to identify the phase space and does not provide us with
specific trajectories on which we integrate to perform a Radon
transform. In fact, not only we will discuss the Radon transform of
functions depending on points on the cylinder (which, to the best of
our knowledge, has never been presented in the literature), but also
intend to study how to construct the map of the positive probability
density distributions living on the phase space onto the family of
the positive probability distributions of random variables living on
the helices. We will address only the classical motion since the
quantized version of the map, that is known for the motion on the
line, needs additional consideration for the motion on the circle,
due to specific properties of compactification in one dimension when
one goes from the plane to the cylinder. The analysis carried out in
this paper might be therefore very relevant for tomography in
quantum mechanics, where we would like to integrate on Lagrangian
submanifolds to have marginals on the transversal Lagrangian leaf
and therefore it becomes relevant for us to understand  what is the
space of all Lagrangian submanifolds and the transversal ones.

The aim of this work is to introduce an invertible tomographic map
of  probability distributions on  phase space of a particle
moving on the circle onto the probability marginal distributions on
the helix of the cylinder (tomograms). The paper is organized as
follows. In Section
\ref{sec-sympltomog} we review the symplectic tomographic approach
for a  free particle moving on the line. Section
\ref{sec-tomogcircle} introduces the tomographic map for functions
on the phase space (cylinder) of the free particle moving on the
circle. We consider an explicit example in Section
\ref{sec-example}. The multidimensional generalization is
considered in Section
\ref{sec-torustomo}. In Section
\ref{sec-limit} we look at the limit of the tomographic map for the
particle moving on the torus when the radii of the circles tend to
infinity and show that in this limit we get the symplectic
tomographic map corresponding to the standard Radon transform.
Perspectives and conclusions are presented in Section
\ref{sec-concl}.

\section{Symplectic tomography}
\label{sec-sympltomog}

Let us consider a function $f(q,p)$ on the phase space $(q,p) \in
\mathbb{R}^2$ of a particle moving on the line $q \in \mathbb{R}$.
The Radon transform as originally formulated solves the following problem: to reconstruct a
function of two variables, say $f(p,q)$, if its integrals over
arbitrary lines are given.

In the $(q,p)$ plane, a line is given by the equation
\begin{equation}
X -\mu q - \nu p = 0.
\end{equation}
By using the homogeneity we may write
\begin{equation}
\tilde{X} - \cos\theta q - \sin\theta p = 0.
\end{equation}
Thus, the family of lines has the manifold structure
$\mathbb{R}\times\mathbb{S}$, with $\mathbb{S}$ the unit circle,
$\tilde{X}\in\mathbb{R}$ and $\theta\in[0,2\pi]$. There is another
way to recover this manifold structure which turns out to be useful
for generalizations to higher dimensions. The Euclidean group
$\mathrm{E}(2)$ acts transitively on the set of lines in the plane
with a stability group given by the translations along the line
itself. Therefore the family of lines is given by
$\mathrm{E}(2)/\mathbb{R}$, i.e.\ $\mathbb{R}\times\mathbb{S}$.

\begin{figure}[t]
\begin{center}
\includegraphics[width=8cm]{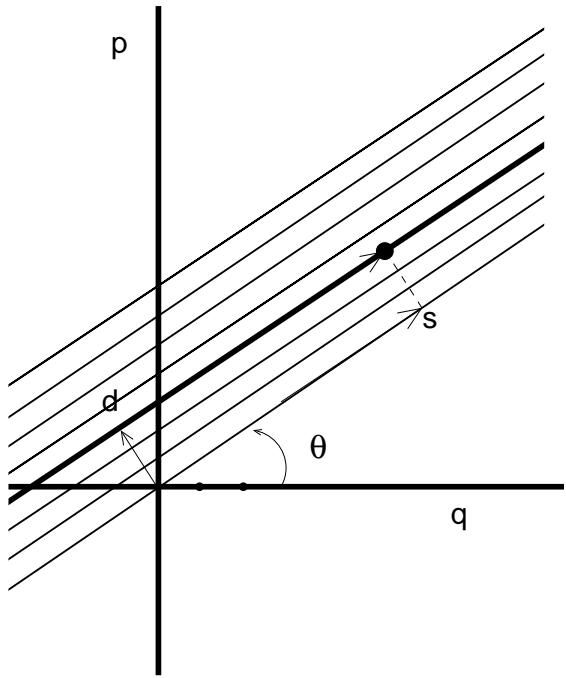}
\end{center}
\caption{Tomography on the plane. }
\label{fig:plane}
\end{figure}

The action of $\mathbb{R}\times\mathbb{S}$ may be visualized in the
following way: a fiducial line passing through the origin may be
translated along the normal to the line to generate a family of
parallel lines. See Fig.~\ref{fig:plane}. Afterwards, by using the
rotation group we may rotate this family of parallel lines into any
other family of parallel lines. As the two actions commute, we may
also rotate first and then translate. Thus, we may consider the set
of all lines passing through the origin and parametrized by the
angle and then translate each one along the normal.

It is interesting to observe that
\begin{equation}
\mathbb{R}^2 = \mathrm{E}(2)/\mathbb{S}, \qquad
\mathbb{R}\times\mathbb{S} = \mathrm{E}(2)/\mathbb{R}.
\end{equation}
The Radon transform maps $\mathcal{F}(\mathbb{R}^2)$ into
$\mathcal{F}(\mathbb{R}\times\mathbb{S})$, where $\mathcal{F}$ is a
suitable class of functions that depends on the physical setting
(for our purposes, $L^1$ is enough). The set of lines can be
parametrized by two numbers: the distance from the origin, $d\in
\mathbb{R}$, and the angle with respect to the $p=0$ axis, $\theta
\in [0,2 \pi)$. Any point in $\mathbb{R}^2$ can be then parametrized
by
\begin{equation}
(q,p)=(s\cos\theta,s\sin\theta)+(-d\sin\theta,d\cos\theta) ,
\end{equation}
where $s$ is the parameter running along the line defined by $d$ and
$\theta$. See Fig.\ \ref{fig:plane}.

The Radon transform is defined by
\begin{equation}
F(d,\theta)=\int_{-\infty}^{+\infty} f(s\cos\theta
-d\sin\theta,s\sin\theta+ d\cos\theta) ds .
\end{equation}
The inversion formula, as given by Radon, amounts to consider first
the average value of $F$ on all lines tangent to the circle of
center $P=(q,p)$ and radius $r$, namely,
\begin{equation}
F_P(r)=\frac{1}{2\pi}\int_0^{2\pi} F(q \cos \theta + p\sin\theta +
r,\theta) d\theta
\end{equation}
and then
\begin{equation}
f(q,p)=-\frac{1}{\pi}\int \frac{F'_P(r)}{r} dr.
\end{equation}
The Radon transform maps a (suitable) function on the plane into a
function on the cylinder. Some conditions that guarantee the
invertibility and continuity of the map were studied by Radon
himself \cite{Rad1917}, John \cite{John}, Helgason \cite{Helgason}
and Strichartz \cite{Strichartz}.

It is possible to write the Radon transform in the affine language
(the so-called tomographic map) \cite{Rad1917,Gelf}
\begin{eqnarray}
\omega_f(X,\mu,\nu) &=& \left\langle \delta(X-\mu q-\nu
p)\right\rangle \nonumber\\
& =& \int_{\mathbb{R}^2} f(q,p) \delta(X-\mu q-\nu p)  dqdp ,
\label{eq:radondef}
\end{eqnarray}
where $\delta$ is the Dirac function and the parameters $X, \mu, \nu
\in \mathbb{R}$. We notice that
\begin{equation}
\left(
  \begin{array}{cc}
    \mu & \nu \\
  \end{array}
\right) \left( \begin{array}{cc}
    1 & 0 \\
    0 & 1 \\
  \end{array}
\right) \left(
  \begin{array}{c}
    q \\
    p \\
  \end{array}
\right) = \mu q + \nu p ,
\end{equation}
but also
\begin{equation}
\left(
  \begin{array}{cc}
    -\nu & \mu \\
  \end{array}
\right) \left(
  \begin{array}{cc}
    0 & -1 \\
    1 & 0 \\
  \end{array}
\right) \left(
  \begin{array}{c}
    q \\
    p \\
  \end{array}
\right) = \mu q + \nu p .
\end{equation}
This means that the argument in the Dirac delta function may be
considered either as a Euclidean product or as a symplectic product.
Equivalently, one might consider the Euclidean or symplectic Fourier
transforms.

Another remark is the following. The full linear inhomogeneous group
acts transitively on the family of lines on $\mathbb{R}^2$. Instead
of $\mathrm{E}(2)$ as a privileged group, we may consider
\begin{equation}
\mathrm{SL}(2,\mathbb{R}) \equiv \mathrm{Sp}(2,\mathbb{R}) \equiv
\mathrm{IGL}(2,\mathbb{R})/(\mathbb{R}^2\times\mathbb{R}),
\end{equation}
where SL, Sp and IGL are the special linear, symplectic and
inhomogeneous linear groups, respectively. The $\mathbb{R}$-group in
the ``denominator'' gives dilations while $\mathbb{R}^2$ gives
translations. Because $\mathrm{Sp}(2,\mathbb{R})$ is not abelian, it
can be generated by two types of transformations: 
rotations
\begin{equation}
\left(
  \begin{array}{cc}
    \cos\theta & \sin\theta \\
    -\sin\theta & \cos\theta \\
  \end{array}
\right),
\end{equation}
and ``squeezing" transformations
\begin{equation}
 \left(
  \begin{array}{cc}
    s & 0 \\
    0 & s^{-1} \\
  \end{array}
\right).
\end{equation}
The action of the ``squeezing" transformation maps lines into lines,
while preserving the area of the triangle. The further action of the
rotation group will change the angle formed with the $q$ axis. One
may show that the Radon transform is equivariant with respect to the
action of $\mathrm{SL}(2,\mathbb{R})$ or $\mathrm{E}(2)$; both of
them preserve the measure on $\mathbb{R}^2$.

The inverse transform of (\ref{eq:radondef}) reads
\cite{Rad1917,Gelf}
\begin{eqnarray}
f(q,p) &=&  \int_{\mathbb{R}^3} \omega_f(X,\mu,\nu) e^{i(X-\mu q-\nu
p)} \frac{dXd\mu d\nu}{(2\pi)^2} . \quad \label{eq:invgen}
\end{eqnarray}
In polar coordinates, $\mu=r\cos \theta$, $\nu= r\sin \theta$,
the inversion formula takes the form of the standard inverse Radon
transform:
\begin{eqnarray}
f(q,p) =  \int_{\mathbb{R}}\int_0^{2\pi} \omega_f(X,\cos
\theta,\sin \theta) K(\theta,q,p) \frac{dXd\theta}{(2\pi)^2},
\label{eq:invradon}
\end{eqnarray}
with
\begin{eqnarray}
K(\theta,q,p)= \sin \theta
\int_0^\infty  e^{-i q r \cos \theta -i p r \sin \theta}
r dr
\label{eq:KerK}
\end{eqnarray}
and where we made use of the homogeneity of $\omega_f(X,\mu,\nu)$
\begin{eqnarray}
\omega_f(\lambda X,\lambda \mu,\lambda \nu) =
\frac{1}{|\lambda|}\omega_f(X,\mu,\nu),
\label{eq:homog}
\end{eqnarray}
that is a direct consequence of (\ref{eq:radondef}). If the function
$f(q,p)$ is a probability density distribution on the phase space of
a classical particle, i.e.
\begin{eqnarray}
f(q,p) \geq 0, \quad \int_{\mathbb{R}^2} f(q,p) dqdp =1,
\label{eq:denscond}
\end{eqnarray}
also the function $\omega_f(X,\mu,\nu)$ is nonnegative and is
called a symplectic tomogram or the ``Radon component" of the
distribution function $f(q,p)$ (analogously to the Fourier
component of a function).  The Radon component contains the same information on
the state of the particle evolving on the phase space as the initial
distribution function. Summarizing:
\begin{eqnarray}
\omega_f(X,\mu,\nu) \geq 0, \quad \int_{\mathbb{R}}
\omega_f(X,\mu,\nu) dX =1, \quad \forall \mu, \nu,
\label{eq:normcond}
\end{eqnarray}
the family of tomograms depends on the two real parameters $\mu$
and $\nu$.

\section{Tomography on the circle}
\label{sec-tomogcircle}

In order to extend the preceding tomographic analysis to particles
confined to compact domains there are two alternative definitions,
following two different strategies.

\subsection{First definition: tomography on the strip}
\label{sec-tomogstrip}

Let us choose for definiteness an interval of width $2\pi$. The
configuration space
\begin{equation}\label{eq:strip}
I=[0,2\pi)
\end{equation}
yields the phase space $I\times\mathbb{R}$ (a strip). To consider
this case it is convenient to deal with the parametrization of
lines given by $\mathbb{S}\times\mathbb{R}$, where $\mathbb{R}$ is
the translation along the normal to the line. If we consider the
intersection of the lines with the selected strip, it is still
possible to consider the treatment of the planar situation, where
in addition the measure $dq dp$ is multiplied by the
characteristic function of the strip.

The state of a classical particle moving in the interval in the
presence of fluctuations is associated with a distribution
function $f(q,p)\geq0$, satisfying the normalization condition
\begin{eqnarray}
\int_{I\times\mathbb{R}} f(q,p) dq dp =1.
\label{eq:normalization1}
\end{eqnarray}
In this case, the symplectic tomogram (\ref{eq:radondef})
specializes to
\begin{eqnarray}
\omega_f(X,\mu,\nu) = \int_{I\times\mathbb{R}} f(q,p) \delta(X-\mu
q-\nu p) dqdp ,
\label{eq:radonstrip}
\end{eqnarray}
with $X,\mu,\nu \in \mathbb{R}$. One easily checks nonnegativity
and normalization like in Eq.\ (\ref{eq:normcond}):
\begin{eqnarray}
\omega_f(X,\mu,\nu) \geq 0, \quad \int_\mathbb{R}
\omega_f(X,\mu,\nu) dX =1, \quad \forall \mu, \nu.
\label{eq:normcond1}
\end{eqnarray}
The inverse transform, still given by (\ref{eq:invgen}), yields a
function
\begin{eqnarray}
f(q,p)=\chi_{I}(q) f(q,p)
\label{eq:fchar}
\end{eqnarray}
($\chi_{I}$ being the characteristic function), that vanishes
identically outside the strip, i.e.\ $f(q,p)=0$ for $q
\notin I$.

On the other hand, a function $f$ on the strip $I\times\mathbb{R}$
can be extended to a periodic function $f_{2\pi}$ over the whole
plane $\mathbb{R}^2$ defined by
\begin{eqnarray}
f_{2\pi}(q,p) &=& \sum_{k\in\mathbb{Z}} f(q-2\pi k,p) \nonumber\\
&=& \sum_{k\in\mathbb{Z}} f(q-2\pi k,p)\chi_{I+ 2\pi k}(q),
\label{eq:perext}
\end{eqnarray}
where the periodicity, $f_{2\pi}(q+2\pi,p)=f_{2\pi}(q,p)$, is
apparent and we used Eq.\ (\ref{eq:fchar}) in the second equality.

\begin{figure}[t]
\begin{center}
\includegraphics[width=8.5cm]{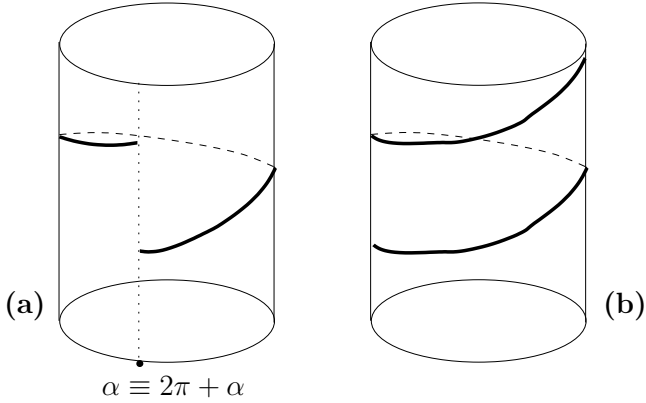}
\end{center}
\caption{Tomography on the circle: (a) strip; (b) cylinder. }
\label{fig:cylinder}
\end{figure}

The phase space has become a cylinder $\mathbb{S}\times\mathbb{R}$,
where $\mathbb{S}=\mathbb{R}/ (2 \pi
\mathbb{Z})$ is the unit circle. In order to emphasize this change
of geometry, we will denote the position of a particle on the circle
by the angle $\phi$ and its angular momentum by $J$. The state of a
classical particle moving on the circle in the presence of
fluctuations is associated with the distribution function
(\ref{eq:perext}) $f(\phi,J)=f_{2\pi} (q=\phi, p=J)$, satisfying the
normalization condition
\begin{eqnarray}
\int_{\mathbb{S}\times\mathbb{R}} d\phi dJ  f(\phi,J) =1.
\label{eq:normalization}
\end{eqnarray}
Due to the periodicity $f(\phi+2k\pi,J)=f(\phi,J)$
($k\in\mathbb{Z}$), in the inversion formula (\ref{eq:invgen}), the
Fourier integral over $\mu$ will be replaced by a Fourier series.
Therefore, it follows that, in order to reconstruct $f(\phi,J)$, in
(\ref{eq:radonstrip}) only the tomograms $\omega_f(X,m,\nu)$ with
$m\in\mathbb{Z}$ are really needed. Thus, we define
\begin{eqnarray}
& &\omega_f^{(0)}(X,m,\nu) = \langle \delta(X-m \phi-\nu J)\rangle
\nonumber\\
& &\quad = \int_{I\times\mathbb{R}} d\phi dJ f(\phi,J) \delta(X-m
\phi-\nu J),
\label{eq:radoncompactdef}
\end{eqnarray}
where $X, \nu \in\mathbb{R}$ and $m\in\mathbb{Z}$. In Eq.\
(\ref{eq:radoncompactdef}) one integrates along the family of
one-step segments of helices: $X = m\phi + \nu J$ with $0 <
\phi < 2\pi$ and $X/\nu -2 m\pi/\nu<J<X/\nu$. The choice of this
family implies the choice of one particular fiber of the cylinder
along which each segment is discontinuous. In fact, observe that
if the $\phi$-domain of integration in (\ref{eq:radoncompactdef})
is changed, say to
\begin{equation}\label{eq:Ialpha}
I_\alpha=I+\alpha=[\alpha,2\pi+\alpha),
\end{equation}
one gets different families of tomograms labeled by a gauge
$\alpha$,
\begin{eqnarray}
\omega_f^{(\alpha)}(X,m,\nu) = \int_{I_\alpha\times\mathbb{R}} d\phi
dJ f(\phi,J),
\label{eq:radoncompactalpha}
\end{eqnarray}
which are related to (\ref{eq:radoncompactdef}) by
\begin{equation}\label{eq:gauge}
\omega_f^{(\alpha)}(X,m,\nu)=\omega_{\tau_\alpha
f}^{(0)}(X-m\alpha,m,\nu) ,
\end{equation}
where $\tau_\alpha f(q,p)=f(q+\alpha,p)$ is a horizontal
translation of $f$. Notice that, due to the periodicity of $f$,
the horizontal tomogram, with $m=0$, is gauge invariant, namely
$\omega_f^{(\alpha)}(X,0,\nu)=\omega_{f}^{(0)}(X,0,\nu)$.
Moreover, all families are obtained by restricting $\alpha\in
[0,2\pi)$. In fact one gets
\begin{equation}\label{eq:alphaperiod}
\omega_f^{(\alpha+2\pi k)}(X,m,\nu)=\omega_{f}^{(\alpha)}(X-2 \pi
m k,m,\nu),
\end{equation}
for  $k\in\mathbb{Z}$. The gauge $\alpha$ is the anomaly of the
chosen fiber of the cylinder $\mathbb{S}\times\mathbb{R}$. See Fig.\
\ref{fig:cylinder}(a). One easily checks nonnegativity and
normalization in the form
\begin{eqnarray}
& & \omega_f^{(\alpha)}(X,m,\nu) \geq 0, \quad \int_\mathbb{R}
\omega_f^{(\alpha)}(X,m,\nu) dX =1, \nonumber \\
& & \qquad \qquad \forall m, \nu, \alpha.
\label{eq:normcond2}
\end{eqnarray}
Let us emphasize again that in these formulas, unlike in Eqs.\
(\ref{eq:normcond}) and (\ref{eq:normcond1}), $m\in\mathbb{Z}$.

The inverse transform is
\begin{equation}
f(\phi,J) = \sum_{m\in\mathbb{Z}} \int_{\mathbb{R}^2}
\omega_f^{(\alpha)}(X,m,\nu) e^{i(X-m \phi-\nu J)} \frac{dX
d\nu}{(2\pi)^2} .
\label{eq:invradoncompactdef}
\end{equation}
Indeed, by making use of the Poisson formula
\begin{equation}\label{eq:Poisson}
\sum_{m\in\mathbb{Z}} e^{i m\phi}= 2\pi \sum_{m\in\mathbb{Z}}
\delta(\phi-2 m\pi) = 2\pi \, \delta_{2\pi}(\phi) ,
\end{equation}
where $\delta_T$ is the $T$-periodic delta function,
\begin{equation}\label{eq:periodicdelta}
\delta_{T}(t)= \delta( t\; (\mathrm{mod}\, T))= \left\{
\begin{array}{ll}
  \sum_{k\in \mathbb{Z}}\delta(t - k T) , & T\neq 0 \\
   & \\
  \delta(t),  & T=0
\end{array}
\right. ,
\end{equation}
one gets
\begin{eqnarray}
& &\sum_{m\in\mathbb{Z}} \int_{\mathbb{R}^2}\frac{dX d\nu}{(2\pi)^2}
e^{i(X-m \phi-\nu J)}
\omega_f^{(\alpha)}(X,m,\nu) \nonumber\\
&=& \int_{I_\alpha\times\mathbb{R}} d\psi dK
f(\psi,K)\sum_{m\in\mathbb{Z}} \int_{
\mathbb{R}^2}\frac{dX d\nu}{(2\pi)^2} \nonumber\\
& & \times e^{i(X-m \phi-\nu J)} \delta(X-m \psi-\nu K)
\nonumber\\
&=& \int_{I_\alpha \times\mathbb{R}} d\psi dK f(\psi,K)
\sum_{m\in\mathbb{Z}}  \frac{e^{i
m(\psi- \phi)}}{2\pi} \delta(K-J) \nonumber\\
&=& \int_{I_\alpha \times\mathbb{R}} d\psi dK f(\psi,K)
\delta_{2\pi}(\psi- \phi) \delta(K-J) \nonumber\\
&=& f(\phi,J),
\end{eqnarray}
as required.

\subsection{Second definition: tomography on the cylinder}
\label{sec-tomogcylinder}

\begin{figure}[t]
\begin{center}
\includegraphics[width=8cm]{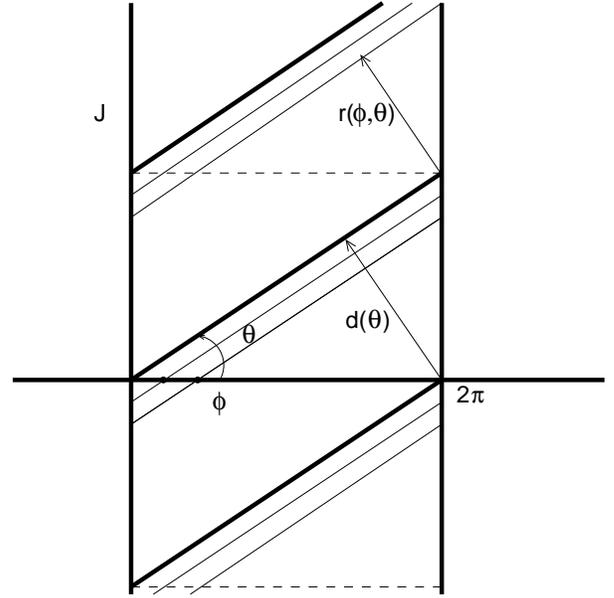}
\end{center}
\caption{Phase space and relevant variables for the tomography on the strip.}
\label{fig:beppe}
\end{figure}

When we restrict our attention to periodic functions, we are
identifying the line at $0$ with the line at $2\pi$. In this way
lines become helices. In this situation, however, a new phenomenon
takes place: translations along the ``normal'' will map the helix
into itself, for translations which are integer multiples of $2 \pi \tan \theta$
{(See Fig.\
\ref{fig:beppe})}.
The set of different helices is, therefore, parametrized by an angle
$\theta\in (-\pi,0)$ and the intercept $\phi\in [0,2\pi)$. Notice
that the value  $\theta=0$ does not correspond to an helix but to an
infinite family of circles ``parallel'' to the base circle. Thus,
the set of helices is a trivial bundle with fiber $\mathbb{S}$ and
base manifold $\mathbb{S}\backslash\{0\}$. where we can use as
coordinates the slope and intercept  $(\theta,\phi)$ or the slope
and the shift with respect to the helix crossing the origin i.e.
$(\theta,r(\phi,\theta))$ with $r(\phi,\theta)=(2 \pi-\phi)
\tan
\theta$ .

Thus, in this setting we would define the Radon transform as going
from functions on $[0,2\pi]\times \mathbb{R}$ to functions on
$\mathbb{S}\times (\mathbb{S}\backslash\{0\})$
.
It seems clear that only specific
applications may suggest to use one or the other. For X-ray
tomography the integration along ``segments'' may be appropriate.
For quantum tomography we may want to integrate along maximal
Lagrangian submanifolds to get the marginals along transversal
Lagrangian submanifolds out of the Wigner function on the full phase
space.

For these reasons we introduce a different tomographic probability
distribution: let
\begin{eqnarray}
& &\tilde\omega_f(X,m,\nu) = \langle \delta_{2\pi m}(X-m \phi-\nu
J)\rangle
\nonumber\\
& & = \int_{\mathbb{S}_1 \times\mathbb{R}} d\phi dJ f(\phi,J)
\delta_{2\pi m}(X-m \phi-\nu J),
\label{eq:radoncompactdef1}
\end{eqnarray}
where $X, \nu \in\mathbb{R}$, $m\in\mathbb{Z}$ and
$\mathbb{S}_{1}=\mathbb{S}$ is the unit circle. Observe that
(\ref{eq:radoncompactdef1}) is independent of the $\phi$-domain of
integration, due to the periodicity of the integrand. By plugging
(\ref{eq:periodicdelta}) into (\ref{eq:radoncompactdef1}) we get for
$m\in\mathbb{Z}\backslash\{0\}$ (and an arbitrary
$\alpha\in\mathbb{R}$)
\begin{eqnarray}
\tilde\omega_f(X,m,\nu) &=&
\int_{I_\alpha\times\mathbb{R}} d\phi dJ \nonumber\\
& & \times f(\phi,J)\sum_{k\in \mathbb{Z}} \delta(X-m \phi-\nu
J-2\pi m k)
\nonumber\\
&=& \sum_{k\in \mathbb{Z}} \int_{(I_\alpha+2\pi
k)\times\mathbb{R}}  d\phi dJ \nonumber\\
& & \times f(\phi-2\pi k ,J) \delta(X-m \phi-\nu J)\nonumber\\
&=&\int_{\mathbb{R}^2}  d\phi dJ  f(\phi,J) \delta(X-m \phi-\nu J)
,
\label{eq:helixint}
\end{eqnarray}
while, for $m=0$,
\begin{eqnarray}
\tilde\omega_f(X,0,\nu) = \int_{\mathbb{S}_1 \times\mathbb{R}}
d\phi dJ  f(\phi,J) \delta(X-\nu J) .
\label{eq:horizhelixint}
\end{eqnarray}
In conclusion, here we integrate over the whole helix, while the
previous Eq.\ (\ref{eq:radoncompactalpha}) was integrated on a
single step of it. Notice also that translations along the line $X-m
\phi-\nu J$ preserve the measure. By using the homogeneity in Eq.\
(\ref{eq:radoncompactdef1}) we may consider the quantity
 $X/m - \phi-(\nu/m) J$ that implies $2\pi m$-periodicity of
$\tilde\omega_f$,
\begin{equation}
\tilde\omega_f(X+2 \pi m,m,\nu) = \tilde\omega_f(X,m,\nu) .
\label{eq:radoncompactdef2}
\end{equation}
Therefore, the tomogram lives on a family of cylinders labeled by
the integer $m$. See Fig.\ \ref{fig:cylinder}(b).

The inverse transform is given by
\begin{equation}
f(\phi,J) = \sum_{m\in\mathbb{Z}} \int_{\mathbb{S}_m \times
\mathbb{R}} \tilde \omega_f(X,m,\nu) e^{i(X-m \phi-\nu J)}
\frac{dX d\nu}{(2\pi)^2} ,
\label{eq:invradoncompactdef1}
\end{equation}
where $\mathbb{S}_{-m}=\mathbb{S}_m = \mathbb{R}/ (2 \pi m
\mathbb{Z})$ is the circle of radius $|m|$ and
$\mathbb{S}_0=\mathbb{R}$ the real line. When $f$ is nonnegative and
normalized as in (\ref{eq:normalization}), one easily obtains
\begin{eqnarray}
\tilde\omega_f(X,m,\nu) \geq 0, \quad \int_{\mathbb{S}_m}
\tilde\omega_f(X,m,\nu) dX =1, \quad \forall m, \nu. \
\label{eq:normcond3}
\end{eqnarray}

The proof of Eq.~(\ref{eq:invradoncompactdef1}) goes as follows
\begin{eqnarray}
& &\sum_{m\in\mathbb{Z}} \int_{\mathbb{S}_m \times
\mathbb{R}}\frac{dX d\nu}{(2\pi)^2} e^{i(X-m \phi-\nu J)}
\tilde \omega_f(X,m,\nu) \nonumber\\
&=& \int_{\mathbb{S}_1 \times\mathbb{R}} d\psi dK
f(\psi,K)\sum_{m\in\mathbb{Z}} \int_{\mathbb{S}_m \times
\mathbb{R}}\frac{dX d\nu}{(2\pi)^2} \nonumber\\
& & \times e^{i(X-m \phi-\nu J)} \delta_{2\pi m}(X-m \psi-\nu K)
\nonumber\\
&=& \int_{\mathbb{S}_1 \times\mathbb{R}} d\psi dK
f(\psi,K)\nonumber\\
& & \times \sum_{m\in\mathbb{Z}} \int_{\mathbb{R}^2} \frac{dX
d\nu}{(2\pi)^2} e^{i(X-m \phi-\nu J)} \delta(X-m \psi-\nu K)
\nonumber\\
&=& \int_{\mathbb{S}_1 \times\mathbb{R}} d\psi dK f(\psi,K)
\sum_{m\in\mathbb{Z}}  \frac{e^{i
m(\psi- \phi)}}{2\pi} \delta(K-J) \nonumber\\
&=& \int_{\mathbb{S}_1 \times\mathbb{R}} d\psi dK f(\psi,K)
\delta_{2\pi}(\psi- \phi) \delta(K-J) \nonumber\\
&=& f(\phi,J),
\label{eq:inversionproof}
\end{eqnarray}
where we made use of Poisson formula (\ref{eq:Poisson}) and of the
equality
\begin{eqnarray}
& & \sum_{m\in\mathbb{Z}} \int_{\mathbb{S}_m} dX e^{i(X-m \phi-\nu
J)} \delta_{2\pi m}(X-m \psi-\nu K)
\nonumber\\
&=& \sum_{m\in\mathbb{Z}} \int_{\mathbb{R}} dX e^{i(X-m \phi-\nu
J)} \delta(X-m \psi-\nu K).
\label{eq:inversionproof1}
\end{eqnarray}
It is easy to see how the transforms (\ref{eq:radoncompactalpha})
and (\ref{eq:radoncompactdef1}) are related. Indeed, for
$m\in\mathbb{Z}\backslash\{0\}$ we get from (\ref{eq:helixint})
\begin{equation}\label{eq:relationm}
\tilde \omega_f(X,m,\nu) = \sum_{r\in\mathbb{Z}}
\omega_f^{(\alpha)}(X-2\pi m r,m,\nu) ,
\end{equation}
while, from (\ref{eq:horizhelixint})
\begin{equation}\label{eq:relation0}
\tilde \omega_f(X,0,\nu) = \omega_f^{(\alpha)}(X,0,\nu) .
\end{equation}
Incidentally, this relation can be used to give an alternative
proof of the inversion formula (\ref{eq:invradoncompactdef1}). In
fact, from the equality
\begin{eqnarray}
& & \int_{\mathbb{S}_m}  \tilde \omega_f(X,m,\nu) e^{i X}dX
\nonumber\\
& = & \sum_{r\in\mathbb{Z}}\int_{[\alpha,\alpha+2\pi m)}
\omega_f^{(\alpha)}(X-2\pi m r,m,\nu) e^{i X}dX
\nonumber\\
& = & \int_{\mathbb{R}} \omega_f(X,m,\nu) e^{i X}dX ,
\label{eq:fourierradon}
\end{eqnarray}
which is trivially valid for $m=0$, the inversion formula
(\ref{eq:invradoncompactdef}) translates into
(\ref{eq:invradoncompactdef1}).

\subsection{A few comments}
\label{sec-comments}

A few comments are in order. If the configuration space is an
interval, the phase space will be a strip and a ``free'' particle
bouncing back and forth will move on a rectangle. If we impose
periodic boundary conditions we get circles parallel to the base.
Clearly, if we want to consider the quantum case, we have to
integrate the Wigner function on Lagrangian subspaces and get the
marginals, out of which we should be able to ``reconstruct" the
function. As we know, we need a ``large'' family of such marginals,
perhaps parametrized by the symplectic group, to be able to
reconstruct the ``state,'' i.e.\ the original Wigner function
\cite{Wig32,Moyal,Hillary84,Marmoopen,MarmoPL}. This viewpoint
differs from the original Radon formulation based on  the set of
geodesic lines of the plane $\mathbb{R}^2$ as Riemannian space (for
the two dimensional case), whereas in our case the relevant lines
are the Lagrangian lines of the symplectic
 plane as  phase space of the one dimensional particle.
In the Radon case the picture is dynamical while in the symplectic
case is purely kinematical.

In our ``classical'' setting, we asked a similar question, i.e.\
how to reconstruct a classical distribution function on phase
space by means of its integrals on a family of one-dimensional
subspaces. In some sense the fact that the family is parametrized
by two numbers appears as a necessary condition for the
reconstruction to be possible.

Finally, it appears that the two ansatz considered in this section
yield two different phase spaces. It is reasonable to expect that
what is a suitable function in one situation, need not be suitable
for the other one. Therefore the two proposals may coexist, once
it is clear that they represent different physical situations. In
general, they will yield different results. In a way, physics will
decide which transform better matches the problem at hand.

\section{Gaussian example}
\label{sec-example}

Let us consider as an illustration the particular case
\begin{eqnarray}
 f(\phi,J) = \frac{1}{(2\pi)^{3/2}} e^{-J^2/2},
\label{eq:exf}
\end{eqnarray}
which is properly normalized, $\int_{\mathbb{S}\times\mathbb{R}} f
=1$. The Radon transform (\ref{eq:radoncompactdef}) yields for
$m\neq0$
\begin{eqnarray}
\omega_f^{(\alpha)}(X,m,\nu) &=& \int_{I_\alpha\times\mathbb{R}}
\frac{d\phi\, d J }{(2\pi)^{3/2}} \, e^{-\frac{J^2}{2}}
\delta(X-m\phi-\nu J)
\nonumber\\
&=& \frac{1}{|\nu|} \int_{I_\alpha} \frac{d\phi}{(2\pi)^{3/2}}
\exp\left(-\frac{(m\phi-X)^2}{2
\nu^2}\right)\nonumber\\
&=& \frac{1}{4\pi|m|}\frac{2}{\sqrt{\pi}}
\int_{\frac{|m|}{\sqrt{2}|\nu|}\left(\alpha-\frac{X}{m}\right)}
^{\frac{|m|}{\sqrt{2}|\nu|}\left(\alpha-\frac{X}{m}+2\pi\right)}
dx\; e^{-x^2}\nonumber\\
&=&
\frac{1}{4\pi|m|}\left[\mathrm{erf}\left(\frac{|m|}{\sqrt{2}|\nu|}\left(\alpha-\frac{X}{m}+2\pi\right)\right)
\right.\nonumber\\
& & \left.   -
\mathrm{erf}\left(\frac{|m|}{\sqrt{2}|\nu|}\left(\alpha-\frac{X}{m}\right)\right)\right],
\label{eq:mneq1}
\end{eqnarray}
where $\mathrm{erf}(x)$ is the error function. On the other hand, if
$m = 0$,
\begin{eqnarray}
\omega_f^{(\alpha)}(X,0,\nu) &=& \int_{I_\alpha\times\mathbb{R}}
\frac{d\phi\, d J }{(2\pi)^{3/2}} \, e^{-\frac{J^2}{2}}
 \delta(X -\nu J) \nonumber \\
&=& \frac{1}{(2\pi)^{1/2} |\nu|}
\exp\left(-\frac{X^2}{2\nu^2}\right).
\label{eq:meq0}
\end{eqnarray}
It is easy to verify that the inverse Radon transform
(\ref{eq:invradoncompactdef}) permits to recover the original
function (\ref{eq:exf}).

On the other hands, the tomograms along the helices read ($m\neq
0$)
\begin{eqnarray}
\tilde\omega_f(X,m,\nu) &=& \int_{\mathbb{S}_1\times\mathbb{R}}
\frac{d\phi\, d J }{(2\pi)^{3/2}} \, e^{-\frac{J^2}{2}}
\delta_{2\pi}(X-m\phi-\nu J)
\nonumber\\
&=& \int_{\mathbb{R}^2} \frac{d\phi\, d J }{(2\pi)^{3/2}} \,
e^{-\frac{J^2}{2}} \delta(X-m\phi-\nu
J)\nonumber\\
&=& \frac{1}{|\nu|} \int_{\mathbb{R}} \frac{d\phi}{(2\pi)^{3/2}}
\exp\left(-\frac{(m\phi-X)^2}{2
\nu^2}\right)\nonumber\\
&=& \frac{1}{2\pi |m|} ,
\label{eq:mneq0}
\end{eqnarray}
while, for $m=0$ it coincides with (\ref{eq:meq0}),
$\tilde\omega_f(X,0,\nu)=\omega_f^{(\alpha)}(X,0,\nu)$. Note that
Eqs.\ (\ref{eq:mneq1}) and (\ref{eq:mneq0}) satisfy
(\ref{eq:relationm}).

It is clear from this example that the two transforms are
different. As we stressed before, both being mathematically
legitimate, a choice should be motivated on physical grounds.

\section{Torus tomography}
\label{sec-torustomo}

The generalization to many particles is straightforward. Let us
consider $N>1$ classical particles, each moving on its own circle.
The system state is described by a probability distribution function
$f(\vec \phi,\vec J) \geq 0$ satisfying the normalization condition
\begin{eqnarray}
\int_{\mathbb{T}^N\times\mathbb{R}^N} d\vec \phi d \vec J f(\vec
\phi , \vec J) =1 ,
\label{eq:normalization2}
\end{eqnarray}
with coordinates $\vec \phi =
(\phi_1,\ldots,\phi_N)\in\mathbb{T}^N=\left(\mathbb{S}\right)^N$ on
the $N-$torus and angular momenta $\vec J =
(J_1,\ldots,J_N)\in\mathbb{R}^N$.

The tomogram of the torus is defined by
\begin{eqnarray}
& &\omega_f^{(0)}(\vec X, \vec m,\vec \nu) = \left\langle
\prod_{k=1}^N
\delta(X_k-m_k \phi_k-\nu_k J_k)\right\rangle \nonumber\\
& & = \int_{I^N\times\mathbb{R}^{N}}\!\!\!\! d\vec\phi d\vec J
f(\vec\phi,\vec J) \prod_{k=1}^N \delta(X_k-m_k \phi_k-\nu_k J_k),
\qquad
\label{eq:torusN}
\end{eqnarray}
with $\vec X, \vec \nu \in\mathbb{R}^N$ and $\vec m\in\mathbb{Z}^N$.
The inverse transform reads
\begin{eqnarray}
f( \vec \phi, \vec J ) &=& \sum_{\vec m \in\mathbb{Z}^N}
\int_{\mathbb{R}^{2N}} \frac{d\vec X d\vec \nu }{(2\pi)^{2N}}
\nonumber\\
& & \quad \times  \omega_f^{(0)}(\vec X, \vec m, \vec \nu)
\prod_{k=1}^N e^{i(X_k-m_k\phi_k-\nu_k J_k)} . \quad
\label{eq:invradonN}
\end{eqnarray}
The tomograms $\omega^{(\vec\alpha)}(\vec X, \vec m,\vec \nu)$ and
$\tilde \omega(\vec X, \vec m,\vec \nu)$ are obtained analogously,
as $N$ dimensional generalizations of (\ref{eq:radoncompactalpha})
and (\ref{eq:radoncompactdef1}).

\section{Limit to the standard Radon transform}
\label{sec-limit}

Let us discuss now how the formulas for the Radon transform (and its
inverse) of a function defined on a cylinder tend to those of the
standard Radon transform of a function defined on the plane in the
limit of infinite radius of the cylinder. To this end, let us first
recall how the Fourier series of a periodic function $f_R(q)$ with
period $R$ and normalization $\int_{-R/2}^{R/2}f_R(q)dq=1$ becomes
the Fourier integral when $R\to\infty$. The Fourier series reads
\begin{equation}\label{F1}
f_R(q)=\sum_{m\in\mathbb{Z}} C_{k_m}e^{-ik_mq},\qquad
k_m=\frac{2\pi}{R}\,m,\qquad (m\in\mathbb{Z})
\end{equation}
and its coefficients are given by
\begin{equation}\label{F2}
 C_{k_m}(R)=\frac{1}{R}\int_{-R/2}^{R/2} f_R(q)e^{i2\pi m q/R}\,dq.
\end{equation}
For $R\to\infty$ the Fourier series becomes the Fourier integral
representation of the function $f(q)=\lim_{R \to \infty} f_R(q)$
defined on the line. Thus Eq.\ (\ref{F1}) becomes
\begin{eqnarray}
 f(q)&=&\lim_{R\to\infty} \sum_{m\in\mathbb{Z}} \Delta k
 \frac{R}{2\pi}\,C_{k_m}e^{-ik_mq}
 \nonumber\\
& = &\int_{-\infty}^\infty C(k)e^{-ikq}\,dk ,
\label{F3}\end{eqnarray}
where $\Delta k= k_{m+1}-k_m=2\pi/R$, and $C(k)=\lim C_{k_m}R/2\pi$.
On the other hand, Eq.\ (\ref{F2}) takes the form
\begin{equation}\label{F4}
 C(k)=\lim_{R\to\infty}\frac{R}{2\pi}\, C_{k_m}=
 \frac{1}{2\pi}\int_{-\infty}^\infty f(q)e^{ikq}\,dq.
\end{equation}

Using these well known limiting relations one can get the limit of
the tomographic map formulae for the particle moving on the circle.
For definiteness we will look at the tomogram
(\ref{eq:radoncompactdef}); the procedure is analogous for the other
tomograms. We first replace Eq.\ (\ref{eq:radoncompactdef}) by a
formula that takes into account the radius $R$ of the circle. Given
a probability density $f(\phi,J)\geq 0$ on the cylinder, by
introducing the new variables $q=\phi R/2\pi$ and $p=J$  and setting
\begin{equation}\label{eq:fR}
f_R(q,p)= \frac{2\pi}{R} f\left(\frac{2\pi q}{R}, p\right),
\end{equation}
we have the tomogram (\ref{eq:radoncompactdef}) in the form
\begin{eqnarray}
 & &\omega_f^{(0)}(X,\mu_m,\nu)=\langle\delta(X-\mu_mq-\nu p)\rangle
 \nonumber\\
 & & = \int_{-R/2}^{R/2} \int_{-\infty}^\infty f_R(q,p)\delta(X-\mu_mq-\nu
 p)\,dq\,dp , \quad
\label{L2}\end{eqnarray}
where $\mu_m=2\pi m/R$ with a  correctly normalized probability
density
\begin{equation}\label{L1}
 \int_{-R/2}^{R/2} \int_{-\infty}^\infty f_R(q,p)\,dq\,dp=1.
\end{equation}
The inverse formula (\ref{eq:invradoncompactdef}) reads
\begin{equation}
f_R(q,p) = \sum_{m\in\mathbb{Z}}\Delta\mu \int_{\mathbb{R}^2}
\omega_f^{(0)}(X,\mu_m,\nu) e^{i(X-\mu_m q-\nu p)} \frac{dX
d\nu}{(2\pi)^2} ,
\label{eq:inv}
\end{equation}
with $\Delta\mu=2\pi/R$. In the limit $R\to\infty$, we get formulae
(\ref{eq:radondef}) and (\ref{eq:invgen}) and the tomographic map on
the circle yields the Radon transform on the plane.

\section{Conclusions and perspectives}
\label{sec-concl}

We have shown that one can map the probability distribution density
$f(\phi,J)$, defined on a cylinder in terms of two random variables
(position $\phi$ and angular momentum $J$), onto a family of
probability distribution densities depending on one random variable
$X$, which is a continuous coordinate on the helix. The family of
helices is labelled by the integer number $m$ and the real number
$\nu$. The map is obtained by means of the Radon transform extended
to the case of a cylinder.

The Radon transform is closely related to the Fourier transform. We
pointed out an important specific property of the Radon transform,
that is valid both for tomographic maps of functions defined on the
plane and on the cylinder: in contrast to the Fourier transform, for
which the Fourier component of the probability density is \emph{not}
a probability density, the Radon component of the probability
density (given on the plane or the cylinder) is again a probability
density and depends on some extra parameters.

We have also straightforwardly extended the Radon transform
construction to the classical motion on a multidimensional torus and
shown that the tomographic map of probability densities on cylinder
becomes the tomographic map of probability density on the plane.
This implies that the two corresponding Radon transforms are related
to each other, in close analogy to the relation between Fourier
series and Fourier integrals for functions on a circle and functions
on a line. One difference should be stressed though: while in the
Fourier case the limit is taken in $L^2$, in the Radon case it is
(obviously) taken in $L^1$. This is apparent in the manipulations of
Sec.\ \ref{sec-limit}.

The quantum extension of the tomographic map for the free motion on
a circle requires additional investigation, due to the well-known
ambiguities in the definition of the analogues of the conjugate
observables angle and angular momentum \cite{hradilangle}.
Similarly, the extension of Radon transforms for curved manifolds in
the present and related contexts deserves additional study
\cite{Helgason}.


\acknowledgments V.I.M.\ was partially supported by Italian INFN and
thanks the Physics Department of the University of Naples for the
kind hospitality. P.F.\ and S.P.\ acknowledge the financial support
of the European Union through the Integrated Project EuroSQIP. The
work of M.A.\ and G.M.\ was partially supported by a cooperation
grant INFN-CICYT. M.A. was also partially supported by the Spanish
CICYT grant FPA2006-2315  and  DGIID-DGA (grant2006-E24/2).


\end{document}